\documentclass[twocolumn,showpacs,aps,prl,superscriptaddress,floatfix]{revtex4}
\usepackage{graphicx}
\usepackage{verbatim}
\usepackage{dcolumn}
\usepackage{amsmath}
\usepackage{epsfig}
\usepackage{subfigure}

\newcommand{\BaBarYear}      {10}
\newcommand{\BaBarNumber}    {012}
\newcommand{\BaBarType}      {PUB}  
\newcommand{\SLACPubNumber}  {14201}

\input babarsym.tex
\def\ra{\rightarrow}
\def\epem{e^+e^-}
\def\kpkm{K^+K^-}
\def\pipm{\pi^+\pi^-}
\def\ksks{K^0_S K^0_S}
\def\kkstar{K^{*\pm} K^\mp}
\def\ppbar{p \overline{p}}

\def\figurebox#1#2#3{%
    \def\arg{#3}%
    \ifx\arg\empty
    {\hfill\vbox{\hsize#2\hrule\hbox to #2{\vrule\hfill\vbox to #1{\hsize#2\vfill}\vrule}\hrule}\hfill}%
    \else
    {\hfill\epsfbox{#3}\hfill}%
    \fi}

\begin{document}

\pagestyle{plain}

\begin{flushleft}
\babar-\BaBarType-\BaBarYear/\BaBarNumber \\
SLAC-PUB-\SLACPubNumber\\
arXiv:1007.3526 [hep-ex]\\
\end{flushleft}

\title{{\large \bf   Search for $f_J(2220)$ in radiative $\jpsi$ decays }}

%
\author{P.~del~Amo~Sanchez}
\author{J.~P.~Lees}
\author{V.~Poireau}
\author{E.~Prencipe}
\author{V.~Tisserand}
\affiliation{Laboratoire d'Annecy-le-Vieux de Physique des Particules (LAPP), Universit\'e de Savoie, CNRS/IN2P3,  F-74941 Annecy-Le-Vieux, France}
\author{J.~Garra~Tico}
\author{E.~Grauges}
\affiliation{Universitat de Barcelona, Facultat de Fisica, Departament ECM, E-08028 Barcelona, Spain }
\author{M.~Martinelli$^{ab}$}
\author{A.~Palano$^{ab}$ }
\author{M.~Pappagallo$^{ab}$ }
\affiliation{INFN Sezione di Bari$^{a}$; Dipartimento di Fisica, Universit\`a di Bari$^{b}$, I-70126 Bari, Italy }
% ...
\collaboration{The \babar\ Collaboration}
\noaffiliation

\date{\today}

\begin{abstract}
We present a search for $f_J(2220)$ production in radiative $\jpsi \rightarrow \gamma f_J(2220)$ decays using 
460 $\rm fb^{-1}$ of data collected with the \babar\ detector at the SLAC PEP-II $e^+e^-$ collider. The $f_J(2220)$ 
is searched for in the decays to $K^+K^-$ and $K^0_S K^0_S$. No evidence of this resonance is observed, and 90\% confidence 
level upper limits on the product of the branching fractions for $\jpsi \rightarrow \gamma f_J(2220)$ and 
$f_J(2220) \rightarrow K^+K^- (K^0_S K^0_S)$ as a function of spin and helicity are set at the level of $10^{-5}$, 
below the central values reported by the Mark III experiment. 
\end{abstract}


\maketitle

\setcounter{footnote}{0}

Evidence for the $f_J(2220)$, a narrow resonance with a mass around $2.2 \gev/c^2$ also known as $\xi(2230)$, was first presented 
by the Mark III Collaboration \cite{Baltrusaitis:1985pu}. The $f_J(2220)$ was seen as a narrow signal above a broad enhancement in 
both $\jpsi \rightarrow \gamma f_J(2220),f_J(2220) \rightarrow \kpkm$ and $\jpsi \rightarrow \gamma f_J(2220), f_J(2220) 
\rightarrow \ksks$ decays. The charged and neutral product branching fractions (PBF) were measured to be $(4.2^{+1.7}_{-1.4} \pm 0.8) 
\times 10^{-5}$ and $(3.1^{+1.6}_{-1.3} \pm 0.7) \times 10^{-5}$ with significance of 3.6 and 4.7 standard deviations, respectively. 
The BES Collaboration has also subsequently reported evidence in radiative $J/\psi$ decays at a comparable level of significance 
\cite{Bai:1996wm}. They reported PBF of $(3.3^{+1.6}_{-1.3} \pm 1.2) \times 10^{-5}$ and $(2.7^{+1.1}_{-0.9} \pm 0.8) \times 10^{-5}$ 
for the $\kpkm$ and $\ksks$ channels. Indications of similar structure produced in $\pi^- p$ and $K^- p$ collisions have been 
seen \cite{Bolonkin:1987hh,Aston:1988yp,Alde:1986nx}, while searches for direct formation in $\ppbar$ collisions 
\cite{Amsler:2001fh,Evangelista:1998zg} or two-photon processes \cite{Benslama:2002pa,Acciarri:2000ex} were inconclusive. 

The unexpectedly narrow width of the $f_J(2220)$, approximately $20 \mev$, triggered speculation about its nature. Besides 
the early hypothesis of a ``light Higgs'' \cite{Haber:1983hw}, conjectures range from a multiquark state, to a hybrid resonance, 
a $\Lambda \overline{\Lambda}$ bound state,  a high-spin $s \overline{s}$ state, or a glueball \cite{Bib:theo}. Intriguingly, lattice 
QCD calculations predict a mass for the ground state tensor $2^{++}$ glueball close to $2.2 \gev/c^2$ \cite{Chen:2004bw,Morningstar:1997f}.

We report herein a search for the $f_J(2220)$ in radiative $\jpsi$ decays, with the $\jpsi$ produced via initial state radiation (ISR) 
in $\epem$ collisions recorded at PEP-II. The emission of ISR allows the study of resonance production over a wide range of $\epem$ 
center-of-mass (CM) energies \cite{Benayoun:1999hm}. The data sample used in this analysis consists of 425 fb$^{-1}$ recorded 
at $\sqrt{s}=10.54 \gev$ and 35 fb$^{-1}$ recorded $40 \mev$ below this energy. With a luminosity-weighted cross section for $\jpsi$ 
production of $35.7$ pb, this dataset contains $16.4 \pm 0.3$ million directly-produced $\jpsi$ decays. 

The \babar\ detector is described in detail elsewhere \cite{Bib:Babar}. Charged particle momenta are measured in a tracking system 
consisting of a five-layer double-sided silicon vertex detector (SVT), and a 40-layer central drift chamber (DCH), immersed 
in a 1.5-T axial magnetic field. Photon and electron energies are measured in a CsI(Tl) electromagnetic calorimeter (EMC). 
Charged particle identification (PID) is performed using an internally reflecting ring-imaging Cherenkov detector and the energy 
loss $dE/dx$, measured by the SVT and DCH. 

Detector acceptance is studied using Monte Carlo (MC) simulation based on GEANT4 \cite{Bib::Geant}. Multiple photon emission from the 
initial-state charged particles is implemented using a structure function technique \cite{Bib::Struct1,Bib::Struct2}. The 
$f_J(2220)$ resonance is modeled by a non-relativistic Breit-Wigner function with a mass of $2.231 \gev/c^2$ and a width of 
$23 \mev$ \cite{pdg}. Several hypotheses for the spin and helicity of the $f_J(2220)$ are considered: spin $J=0$, and 
spin $J=2$ with pure helicity $\pm 2$, $\pm 1$ or $0$. The hypothesis $J=4$ is strongly disfavored by lattice QCD 
calculations \cite{Liu:2001wqa}. 

The $\jpsi \ra \gamma \kpkm$ decay is reconstructed by combining two oppositely charged tracks, identified as kaons, with 
a photon candidate. Events containing a $\pi^0$ candidate, defined as a pair of photons of energy larger than 
$50 \mev$ \cite{Edef} having an invariant mass in the range $115 - 155 \mev/c^2$, are discarded. The contamination of 
$\jpsi \rightarrow K^{*\pm}(892)(K^{\pm}\pi^{0})K^{\mp}$, in which the $\pi^0$ is not reconstructed, is further 
reduced by rejecting $\jpsi$ candidates having a kaon with a momentum larger than $1.35 \gev/c$ in the $\jpsi$ CM frame. 

The $\jpsi \rightarrow \gamma \ksks$ channel, examined in $\jpsi \ra \gamma \pipm \pipm$, is reconstructed using events containing 
a photon and four charged tracks. Neutral kaon candidates are reconstructed from $ K^0_S \rightarrow \pipm$, combining a pair of 
oppositely charged tracks identified as pions, with an invariant mass in the range $| M_{\pipm} -M_{K_s} |< 15 \mev/c^2$. To improve 
the signal purity, the angle in the transverse plane between the momentum and the flight direction of each kaon is required to be 
less than 0.1 rad. No $\pi^0$ veto is applied, as the $\jpsi \ra \ksks \pi^0$ decay is forbidden by $C$-parity conservation and 
the overall $\pi^0$ contamination is negligible.

Events with additional charged tracks are rejected. The photon emitted by the $\jpsi$ is also required to have an energy larger than 
$300 \mev$ to suppress background from additional ISR photons or noise from the calorimeter. Finally, the helicity angle of each kaon, 
$\zeta_{K}$, must satisfy $|\cos\zeta_{K}|<0.7$. 

\begin{figure}[tbh]
\begin{center}
\includegraphics[width=0.23\textwidth]{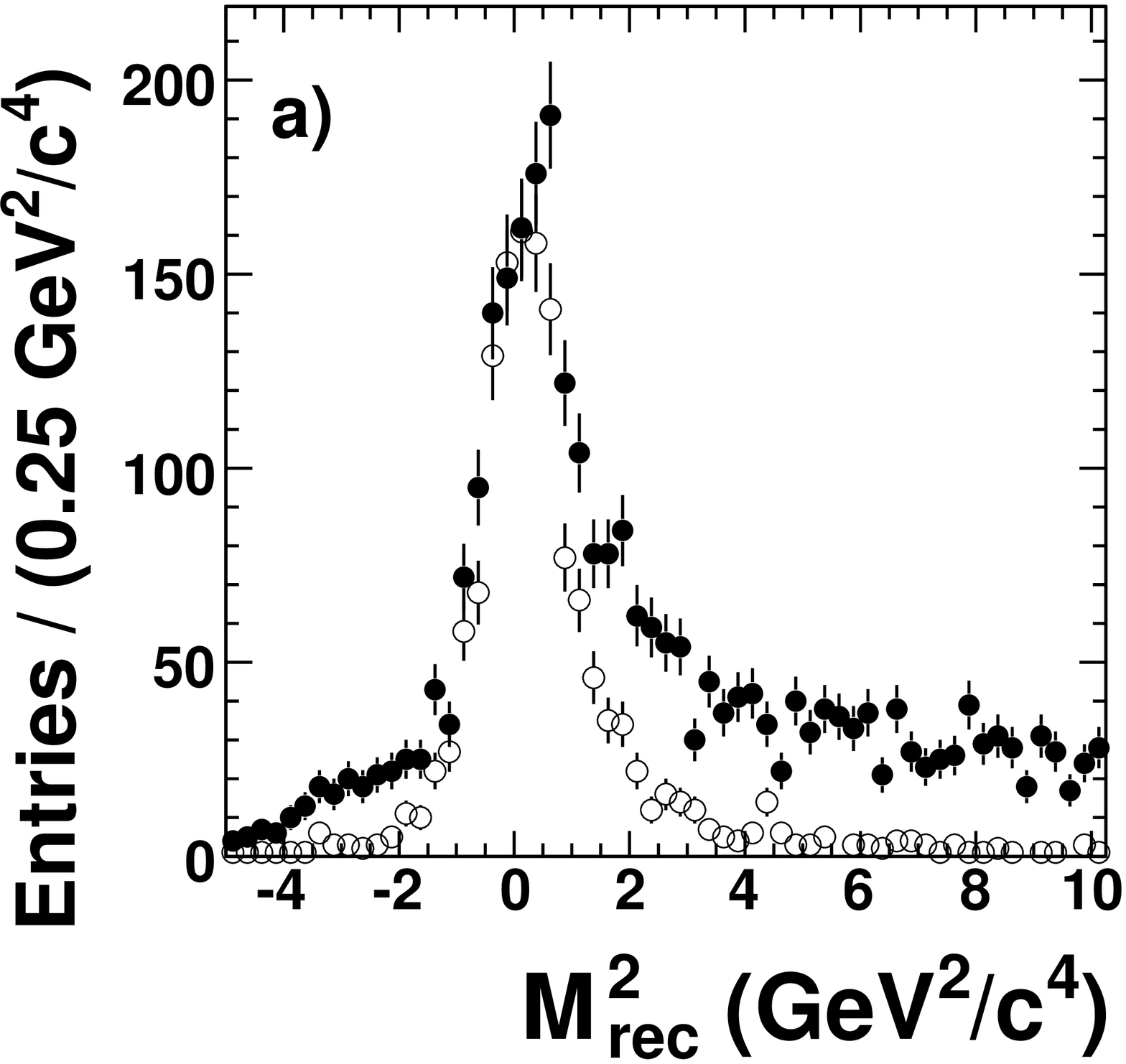}
\includegraphics[width=0.23\textwidth]{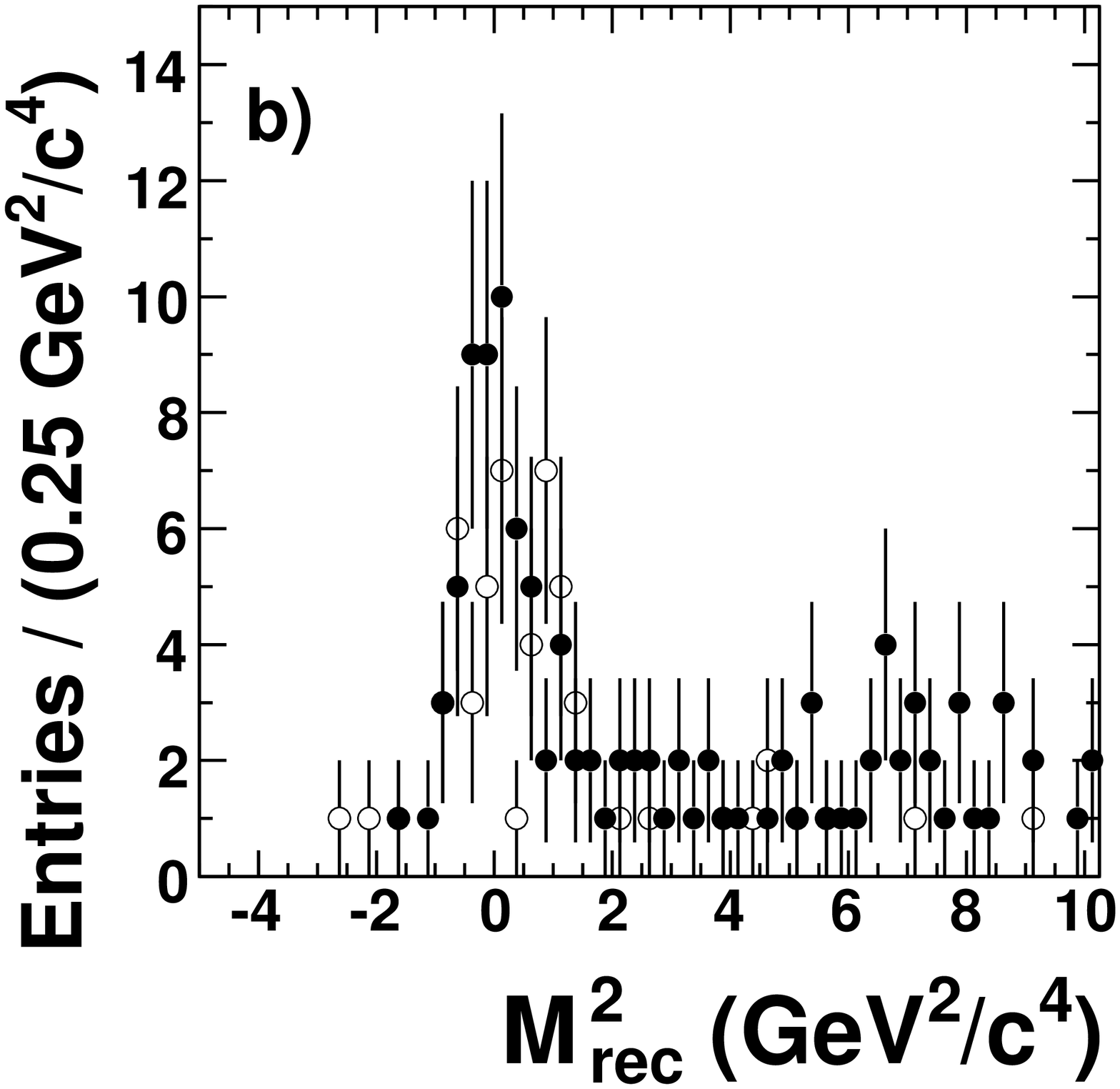}
\caption{The distribution of $M_{rec}^2$, the square of the recoiling mass against the $\jpsi \rightarrow \gamma \kpkm$ (a) 
and $\jpsi \rightarrow \gamma \ksks$ (b) candidates, after all other selection criteria are applied for events in which 
the ISR photon is detected (open circle) or undetected (solid circle).}
\label{fig:mrec}
\end{center}
\end{figure}

Radiative $\epem \rightarrow \gamma_{ISR} \jpsi$ events are then identified. Clusters in the EMC not associated with 
charged-particle tracks and having energy larger than $1 \gev$ are taken as ISR photon candidates. Events in which the 
ISR photon falls within the detector acceptance are selected by demanding an angle between the $\jpsi$ candidate and the 
ISR photon in the CM frame larger than 3.12 (3.10) rad for the charged (neutral) mode. In the opposite case, the square 
of the mass recoiling against the $\jpsi$ is required to lie between $-2.0\gev^2/c^4$ ($-2.0\gev^2/c^4$) and 
$2.0\gev^2/c^4$ ($5.0\gev^2/c^4$) for $\jpsi \rightarrow \gamma \kpkm (\ksks)$ candidates. In both cases, no additional 
photons with energy exceeding $300 \mev$ can be present. For the charged mode, the cosine of the polar angle 
of the photon emitted by the $\jpsi$ is required to be less than 0.8, and, for events where the ISR photon is undetected, 
that of each kaon must be less than 0.9. The distribution of the recoiling mass squared after applying all other 
cuts is displayed in Fig.~\ref{fig:mrec} for combinations having a mass in the range 
$2.8 < m_{\gamma KK} < 3.4 \gev/c^2$. Clear peaks corresponding to ISR events are visible. 

\begin{figure}[!htb]
\begin{center}
\includegraphics[width=0.23\textwidth]{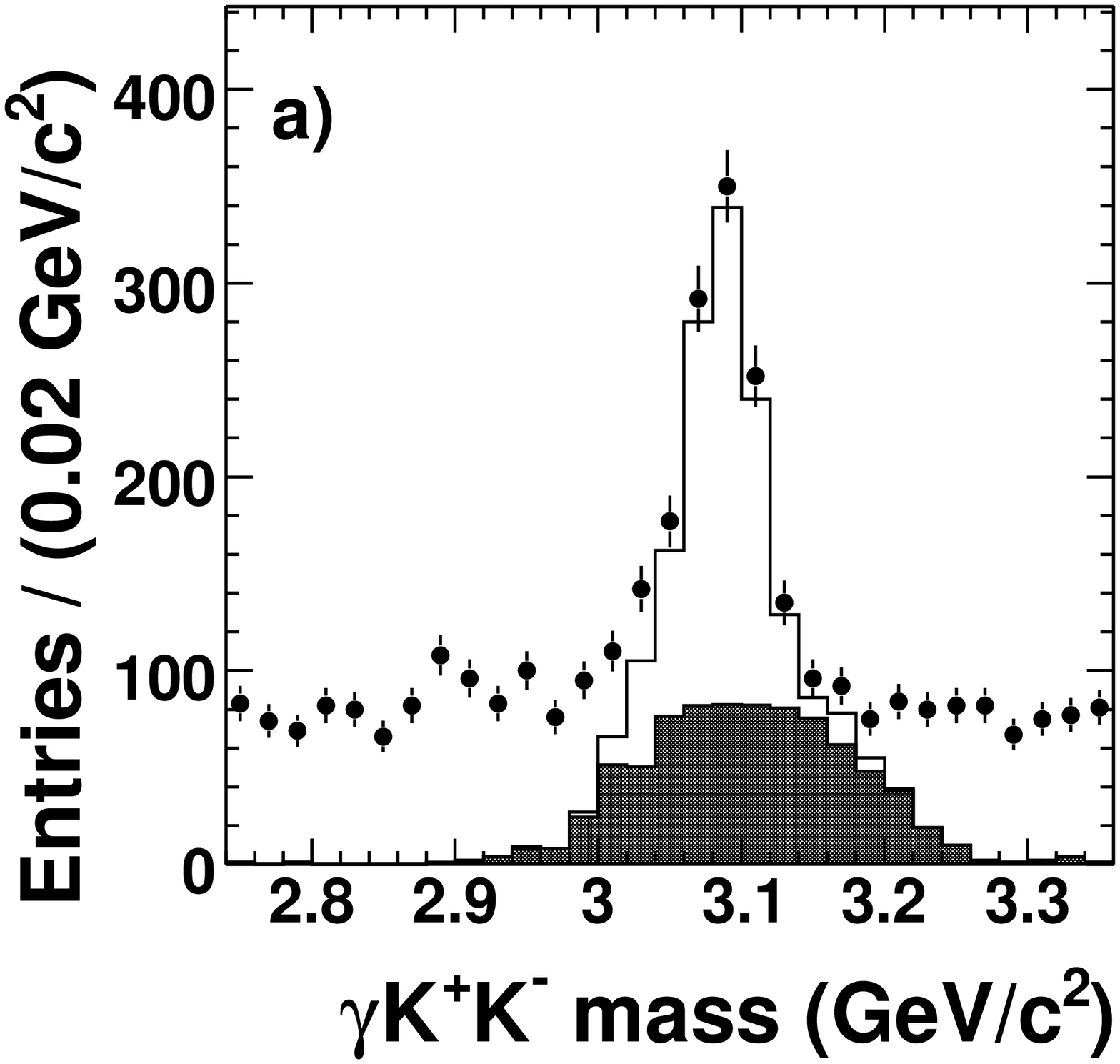}
\includegraphics[width=0.23\textwidth]{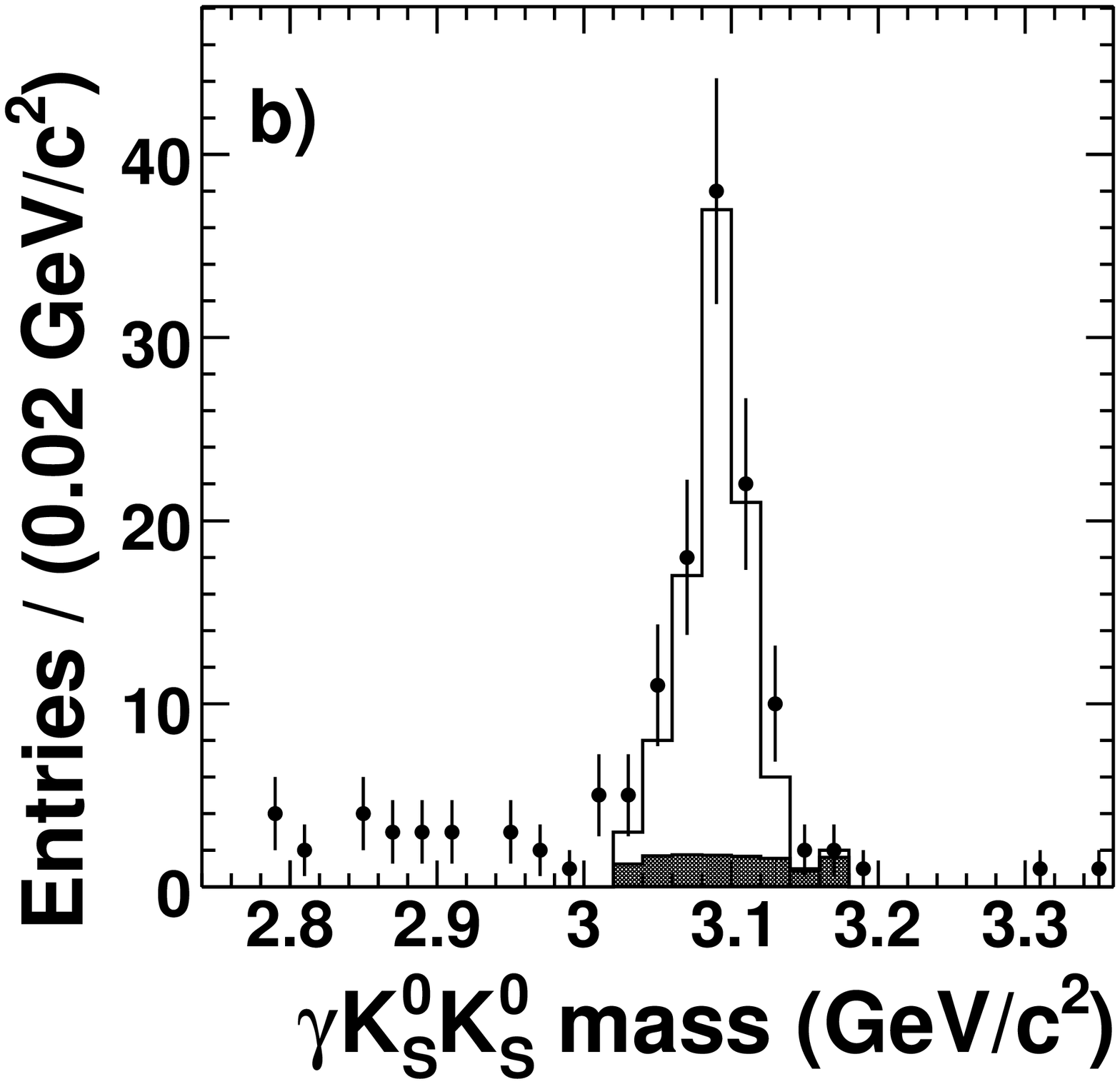}
\caption{The $ \gamma\kpkm$ (a) and $\gamma\ksks$ (b) mass spectra after all selection criteria are applied. The points 
represent data, and the plain histograms show combinations having fit probability larger than 0.01. The estimated inclusive 
background in the final sample is shown as a filled histogram.}
\label{fig:jpsi}
\end{center}
\end{figure}

The resulting $\gamma\kpkm$ and $\gamma\ksks$ mass distributions are displayed in Fig.~\ref{fig:jpsi}. A large 
$\jpsi$ signal over a smooth background is observed for both channels. This background, hereafter referred to as 
inclusive, arises mainly from partially-reconstructed $\jpsi \rightarrow KK + X$ decays and 
$\epem \rightarrow q\overline{q} \gamma_{ISR} \,(q=u,d,s,c)$ production. Its level in the $\jpsi$ region 
is determined by fitting the data with a Gaussian and a second (first) order polynomial 
for the charged (neutral) mode.
The $\jpsi$ candidates are then fitted, constraining their mass to the world-average value \cite{pdg} and requiring a 
common vertex for the decay products. A mass constraint on both $K^0_S$ candidates is also imposed for the neutral 
channel. Combinations having a fit probability larger than 0.01 are retained to form the final sample. The 
corresponding inclusive background is evaluated by correcting the values extrapolated from the unconstrained 
mass spectra for the efficiency of the fit probability cut. 

\begin{figure}[!htb]
\begin{center}
\includegraphics[width=0.38\textwidth]{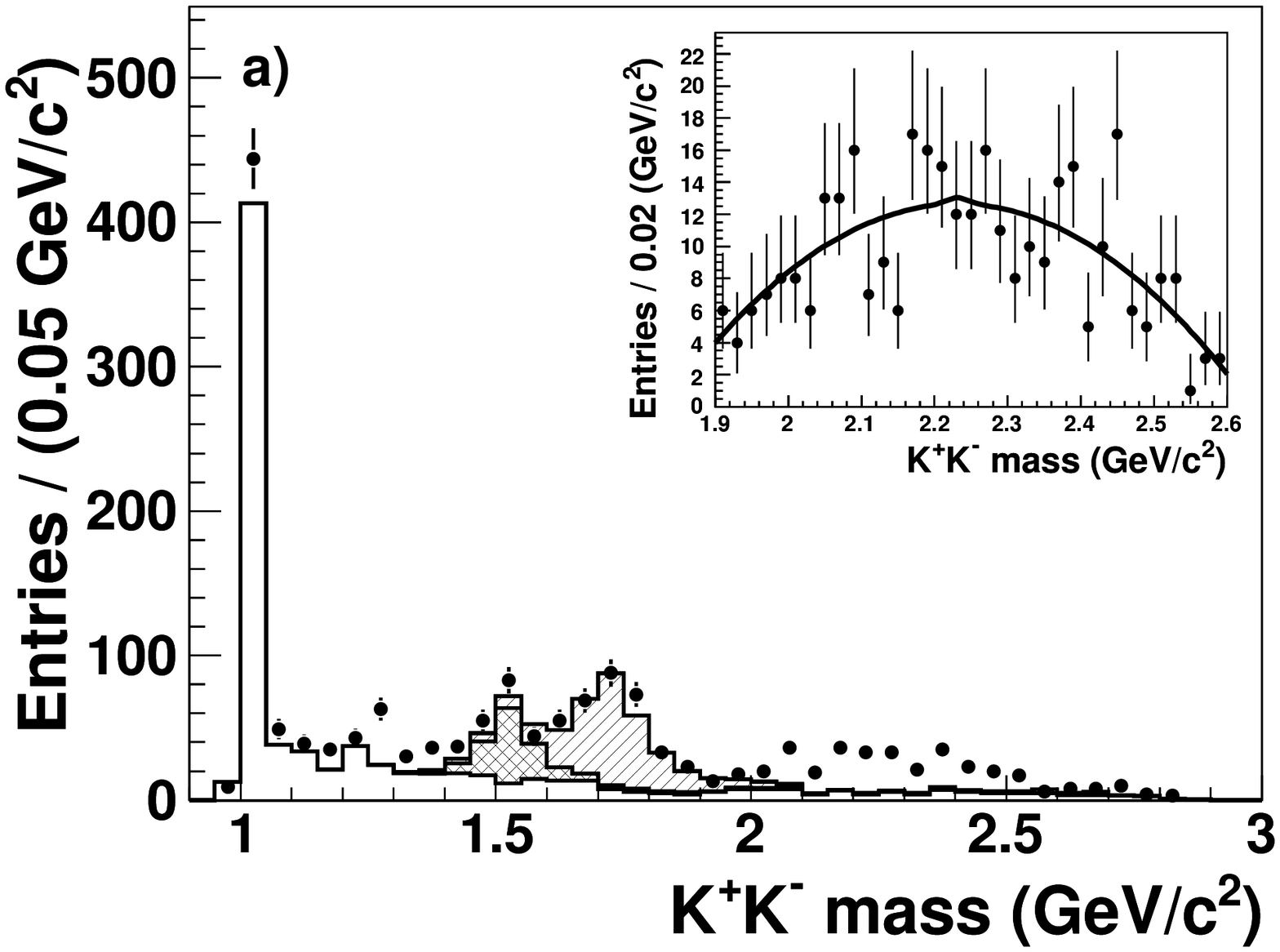}\\
\includegraphics[width=0.38\textwidth]{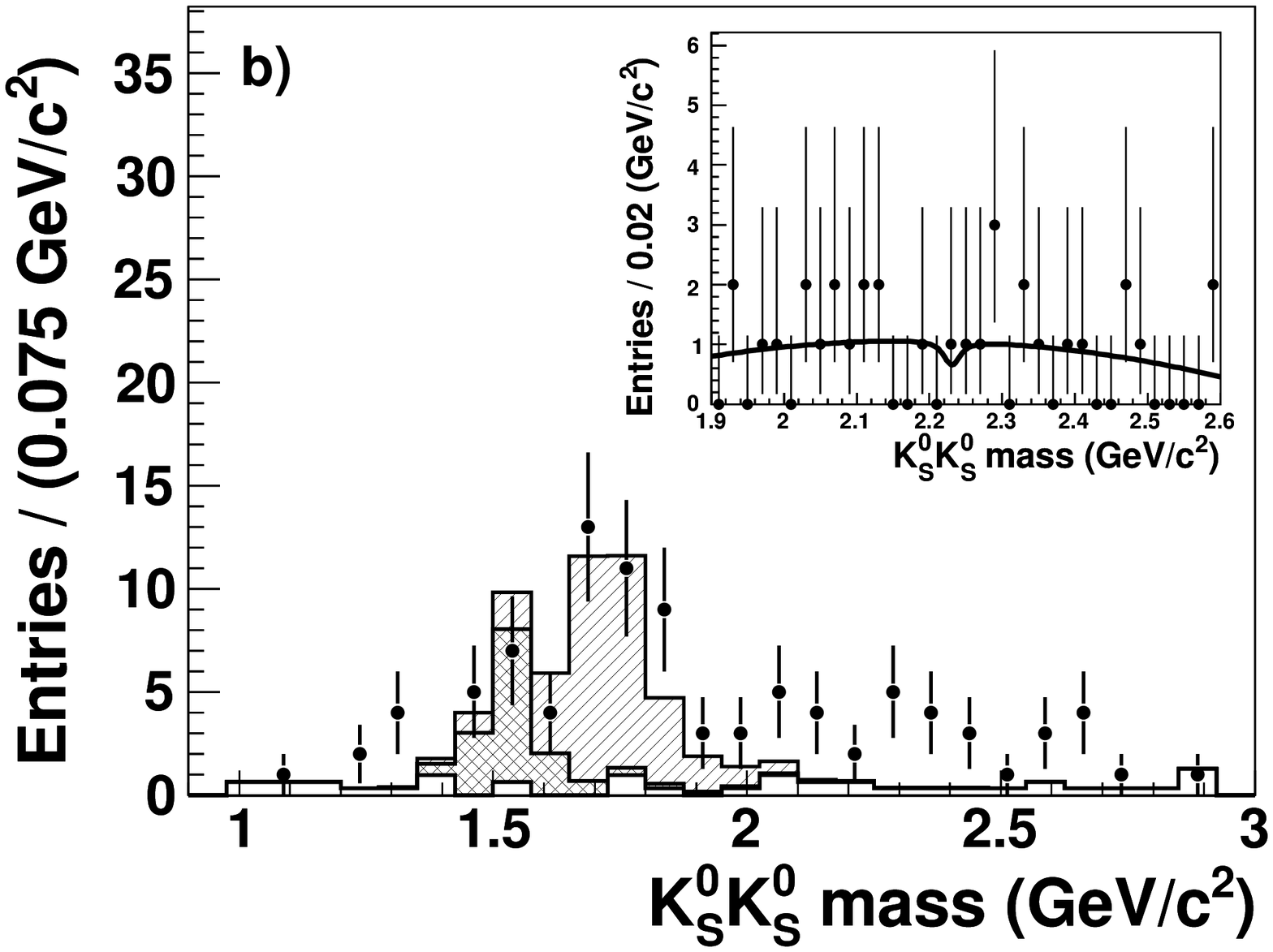}
\caption{The fitted $\kpkm$ (a) and $\ksks$ (b) mass spectra. The expected contributions of the inclusive background 
(plain histogram), $\jpsi \rightarrow \gamma f'_2(1525)$ (cross-hatched histogram), $\jpsi \rightarrow \gamma 
f_0(1710)$ (hatched histogram) are also shown. The results of the fits are displayed in the inserts.}
\label{fig:xi}
\end{center}
\end{figure}

The fitted $\kpkm$ and $\ksks$ mass spectra are shown in Fig.~\ref{fig:xi}, together with the contribution
of  various $\jpsi$ decays and the inclusive background. The shape of the inclusive background is modeled using 
sideband data taken from the unconstrained mass spectra in the ranges $2.7 < m_{\gamma KK} < 2.9 \gev/c^2$ and 
$3.2 < m_{\gamma KK}<3.4 \gev/c^2$. The contributions of the $\jpsi \ra \gamma f'_2(1525)$, 
$\jpsi \ra \gamma f_0(1710)$, and $\jpsi \ra \kkstar$ channels are estimated from MC simulation using world-average 
branching fractions \cite{pdg}. Contamination from $\jpsi \ra \kkstar$ decays is found to be negligible. 
The $f'_2(1525)\rightarrow \kpkm$ and $f'_2(1525)\rightarrow \ksks$ decays are modeled using helicity amplitude 
ratios $x^2= 1.0$ and $ y^2 = 0.44$ \cite{Bai:2003ww}. No interference between the $f_0(1710)$ and 
the inclusive background is considered. The sum of these components accounts for most of the data in the region 
below $2 \gev/c^2$ and reproduces well the contribution of $\phi(1020)$ mesons. The excess seen around $1.25-1.30 \gev/c^2$ 
in the charged mode is likely due to $\jpsi \rightarrow \rho^0 \pi^0, \rho^0 \rightarrow \pipm$ decays, where both 
charged pions are misidentified as kaons, and a photon from the $\pi^0$ decay goes undetected. The data above 
$2 \gev/c^2$ are dominated by partially-reconstructed $\jpsi$ decays.

The number of signal events is determined using an unbinned maximum likelihood fit in the range 
$1.9 \gev/c^2 < m_{KK} < 2.6 \gev/c^2$. The signal is described by a Breit-Wigner distribution convolved with a 
Gaussian resolution function, while the background is modeled by a second order Chebychev polynomial. The mass and 
width of the resonance are fixed to $2.231 \gev/c^2$ and $23 \mev$, respectively. The Gaussian resolution, taken 
from MC simulations, is set to $8 \mev/c^2$ ($6 \mev/c^2$) for the $\kpkm$ ($\ksks$) channel. We have checked on 
a number of independent control samples that the two-body invariant mass resolution is well reproduced by the MC
over the whole invariant mass range studied in this paper. The results of the fits are displayed in 
Fig.~\ref{fig:xi}. No evidence of a $f_J(2220)$ signal is observed.

\begin{table}[htb]
\caption{The efficiency, the PBF of the decays $\jpsi \rightarrow \gamma f_J(2220),$ $f_J(2220) \rightarrow \kpkm$ 
and $\jpsi \rightarrow \gamma f_J(2220), f_J(2220) \rightarrow \ksks$ and 
corresponding 90\% confidence level upper limit (UL) as a function of the spin $J$ and helicity $h$ assumed for 
the $f_J(2220)$. The number of $f_J(2220) \rightarrow \kpkm (\ksks)$ events determined from the fit 
is $1.0^{+8.9}_{-7.9} \pm 1.5$ ($-0.8^{+2.1}_{-1.2} \pm 0.6$). The first uncertainty is statistical 
and the second systematic.}
\begin{center}
\begin{tabular}{l|c | c | c }\hline\hline
Spin / helicity            &    Efficiency       & PBF                               & UL         \\
hypothesis                 &    (\%)             & ($\times 10^{-5}$)                & ($\times 10^{-5}$)  \\\hline
\noalign{\vskip1pt}\multicolumn{4}{c}{$f_J(2220) \rightarrow \kpkm$}                          \\\hline
\noalign{\vskip1pt}
$J=0$                      &  $5.15\pm 0.03$  &  $0.12^{+1.05}_{-0.94} \pm 0.17$  &  $<1.9$    \\
$J=2$ / $h=0$              &  $2.74\pm 0.04$  &  $0.22^{+1.97}_{-1.76} \pm 0.33$  &  $<3.6$    \\
$J=2$ / $h=\pm 1\,$        &  $5.22\pm 0.05$  &  $0.12^{+1.03}_{-0.93} \pm 0.17$  &  $<1.9$    \\
$J=2$ / $h=\pm 2\,$        &  $6.69\pm 0.05$  &  $0.09^{+0.81}_{-0.72} \pm 0.13$  &  $<1.5$    \\ \noalign{\vskip1pt} \hline
\noalign{\vskip1pt}\multicolumn{4}{c}{$f_J(2220) \rightarrow \ksks$}                           \\\hline
\noalign{\vskip1pt}
$J=0$                      &  $1.32\pm 0.01$  &  $-0.39^{+0.96}_{-0.56} \pm 0.28$  &  $<1.7$   \\
$J=2$ / $h=0$              &  $0.74\pm 0.01$  &  $-0.69^{+1.71}_{-1.00} \pm 0.49$  &  $<2.9$   \\
$J=2$ / $h=\pm 1\,$        &  $1.39\pm 0.02$  &  $-0.37^{+0.92}_{-0.54} \pm 0.26$  &  $<1.6$   \\
$J=2$ / $h=\pm 2\,$        &  $1.75\pm 0.02$  &  $-0.29^{+0.73}_{-0.42} \pm 0.21$  &  $<1.2$   \\\noalign{\vskip1pt}
\hline\hline
\end{tabular}
\label{tab:res}
\end{center}
\end{table}

The largest sources of systematic uncertainty arise from the parametrization of the signal and background shapes. 
An uncertainty of 0.2 events arises from fixing the mass, width and resolution of the signal in each channel. This 
contribution is estimated by varying each parameter by $\pm 1 \sigma$ in the fitting procedure. Similarly, 
the uncertainty due to the background parametrization, evaluated to be 1.4 (0.6) events for the $\kpkm$ ($\ksks$) 
mode, is assessed by repeating the fit with a third order Chebychev polynomial. Multiplicative systematic uncertainties on the 
charged (neutral) PBF include the selection procedure [4.0\% (2.2\%)], the determination of the number of $\jpsi$ 
mesons [3.0\% (3.0\%)], the trigger efficiencies [3.1\% (3.5\%)], the track and neutral cluster reconstruction 
[1.9\% (3.3\%)], the particle identification [1.4\% (-)], and the MC statistics [1.0\% (1.4\%)].  

The $\jpsi \rightarrow \gamma f_J(2220), f_J(2220) \rightarrow \kpkm$ and $\jpsi \rightarrow \gamma f_J(2220), f_J(2220) 
\rightarrow \ksks$ PBF are given in Table~\ref{tab:res} as a function of the spin and helicity 
assumed for the $f_J(2220)$. The efficiencies are determined from the corresponding MC and include the 
$K^0_S \rightarrow \pipm$ branching fraction as well as corrections for particle identification, photon detection and 
$K^0_S$ reconstruction. The 90\% confidence level (CL) Bayesian upper limits, based on priors uniform in branching fraction 
and including systematic uncertainties, are also shown. 

In conclusion, no evidence is observed for the $f_J(2220)$ in radiative $\jpsi$ decay in ISR events produced in $\epem$ 
collisions at $\sqrt{s}=m_{\Upsilon(4S)}$. For all hypotheses of spin and helicity, the 90\% CL upper limits on the 
$\jpsi \rightarrow \gamma f_J(2220), f_J(2220) \rightarrow \kpkm$ and $\jpsi\rightarrow \gamma f_J(2220), f_J(2220) 
\rightarrow \ksks $ PBF are below the central values reported by Mark III. Only one hypothesis of spin and helicity 
($J=2$ and $h=0$) is compatible with the BES results for both final states, while all other possibilities are clearly 
excluded.

We are grateful for the excellent luminosity and machine conditions
provided by our \pep2\ colleagues, 
and for the substantial dedicated effort from
the computing organizations that support \babar.
The collaborating institutions wish to thank 
SLAC for its support and kind hospitality. 
This work is supported by
DOE
and NSF (USA),
NSERC (Canada),
CEA and
CNRS-IN2P3
(France),
BMBF and DFG
(Germany),
INFN (Italy),
FOM (The Netherlands),
NFR (Norway),
MES (Russia),
MICIIN (Spain),
STFC (United Kingdom). 
Individuals have received support from the
Marie Curie EIF (European Union),
the A.~P.~Sloan Foundation (USA)
and the Binational Science Foundation (USA-Israel).


\begin{thebibliography}{99}

\bibitem{Baltrusaitis:1985pu}
  R.~M.~Baltrusaitis {\it et al.} (Mark III Collab.), Phys.\ Rev.\ Lett.\  {\bf 56}, 107 (1986).

\bibitem{Bai:1996wm}
  J.~Z.~Bai {\it et al.} (BES Collab.), Phys.\ Rev.\ Lett.\  {\bf 76}, 3502 (1996).

\bibitem{Bolonkin:1987hh}
  B.~V.~Bolonkin {\it et al.}, Yad.\ Fiz.\  {\bf 46}  799 (1987) [Nucl.\ Phys.\  B {\bf 309} 426 (1988)].

\bibitem{Aston:1988yp}
  D.~Aston {\it et al.}, Phys.\ Lett.\  B {\bf 215} 199 (1988).

\bibitem{Alde:1986nx}
  D.~Alde {\it et al.}, Phys.\ Lett.\  B {\bf 177}  120 (1986).

\bibitem{Amsler:2001fh}
  C.~Amsler {\it et al.} (Crystal Ball Collab.), Phys.\ Lett.\  B {\bf 520}, 175 (2001).

\bibitem{Evangelista:1998zg}
  C.~Evangelista {\it et al.} (JETSET Collab.), Phys.\ Rev.\  D {\bf 57}, 5370 (1998) and references therein.

\bibitem{Benslama:2002pa}
  K.~Benslama {\it et al.}  (CLEO Collab.),  Phys.\ Rev.\  D {\bf 66}, 077101 (2002).
  
\bibitem{Acciarri:2000ex}
  M.~Acciarri {\it et al.}  (L3 Collab.),  Phys.\ Lett.\  B {\bf 501}, 173 (2001).

\bibitem{Haber:1983hw}
  R.~M.~Barnett, G.~Senjanovic and D.~Wyler, Phys.\ Rev.\  D {\bf 30}, 1529 (1984) 
  and references therein.

\bibitem{Bib:theo}
  M.~S.~Chanowitz and S.~R.~Sharpe, Phys.\ Lett.\  B {\bf 132}, 413 (1983);
  K.-T.~Chao, Commun.\ Theor.\ Phys.\  {\bf 3}, 757 (1984);
  S.~Pakvasa, M.~Suzuki, and S.~F.~Tuan, Phys.\ Lett.\  B {\bf 145}, 135 (1984);
  M.~P.~Shatz, Phys.\ Lett.\  B {\bf 138}, 209 (1984);
  A.~Le Yaouanc {\it et al.}, Z.\ Phys.\  C {\bf 28}, 309 (1985);
  S.~Godfrey, R.~Kokoski and N.~Isgur, Phys.\ Lett.\  B {\bf 141}, 439 (1984);
  B.~F.~L.~Ward, Phys.\ Rev.\  D {\bf 31}, 2849 (1985) [Erratum-ibid.\  D {\bf 32}, 1260 (1985)].
  S.~Ono, Phys.\ Rev.\  D {\bf 35}, 944 (1987);
  K.-T.~Chao, Phys.\ Rev.\ Lett.\  {\bf 60}, 2579 (1988); 

\bibitem{Chen:2004bw}
  J.-X.~ Chen and J.-C.~Su, Phys.\ Rev.\  D {\bf 69}, 076003 (2004).

\bibitem{Morningstar:1997f}
  C.~J.~Morningstar and M.~J.~Peardon, Phys.\ Rev.\  D {\bf 56}, 4043 (1997).

\bibitem{Benayoun:1999hm} 
  See for example M.~Benayoun {\it et al.}, Mod. Phys. Lett.  A {\bf 14}, 2605 (1999).
  
\bibitem{Bib:Babar} 
  B.~Aubert {\it et al.} (\babar\ Collab.), Nucl.\ Instrum.\ Meth.\ A {\bf 479}, 1 (2002).

\bibitem{Bib::Geant}  
  S.~Agostinelli {\it et al.} (GEANT4 Collab.), Nucl.\ Instrum.\ Meth.\  A {\bf 506}, 250 (2003).

\bibitem{Bib::Struct1} 
  A.~B.~Arbuzov {\em et al.}, J. High Energy Phys. {\bf 9710}, 001 (1997).

\bibitem{Bib::Struct2} 
  M.~Caffo, H.~Czy\.z, and E.~Remiddi, Nuovo Cim. {\bf A110}, 515  (1997); 
  Phys. Lett. {\bf B327}, 369 (1994).

\bibitem{pdg} 
  C. Amsler et al. (Particle Data Group), Phys. Lett. B {\bf 667}, 1 (2008) and 2009 update.

\bibitem{Liu:2001wqa}
  D.~Q.~Liu and J.~M.~Wu,  Mod.\ Phys.\ Lett.\  A {\bf 17}, 1419 (2002).

\bibitem{Edef} All kinematic quantities are defined in the laboratory frame unless another frame is specified.

\bibitem{Bai:2003ww}
  J.~Z.~Bai {\it et al.} (BES Collab.), Phys.\ Rev.\  D {\bf 68}, 052003 (2003).

\end{thebibliography}
\end{document}